\documentclass[showpacs,amssymb,aps,twocolumn]{revtex4}
\usepackage{amsmath}
\usepackage{amstext}
\usepackage{amsopn}
\usepackage{amsfonts}
\usepackage{amssymb}
\usepackage{bbm}
\usepackage{accents}
\usepackage{empheq}%for overbracket
\usepackage{graphicx}
\usepackage{epsf}
\usepackage{graphics}
\usepackage[latin1]{inputenc}

\begin{document}

\title{Finite Temperature Effective Actions}

\author{Ashok Das$^{a,b}$ and J. Frenkel$^{c}$\footnote{$\ $ e-mail: das@pas.rochester.edu,  jfrenkel@fma.if.usp.br}}
\affiliation{$^a$ Department of Physics and Astronomy, University of Rochester, Rochester, NY 14627-0171, USA}
\affiliation{$^b$ Saha Institute of Nuclear Physics, 1/AF Bidhannagar, Calcutta 700064, India}
\affiliation{$^{c}$ Instituto de Física, Universidade de São Paulo, 05508-090, São Paulo, SP, BRAZIL}

\begin{abstract}
We present, from first principles, a direct method for evaluating the exact fermion propagator in the presence of a general background field at finite temperature, which can be used to determine the finite temperature effective action for the system. As applications, we determine the complete one loop finite temperature effective actions for $0+1$ dimensional QED as well as the Schwinger model. These effective actions, which are derived in the real time (closed time path) formalism, generate systematically all the Feynman amplitudes calculated in thermal perturbation theory and also show that the retarded (advanced)  amplitudes vanish in these theories.
\end{abstract}


\maketitle
\newpage

The effective action for a system of fermions interacting with a background field, which incorporates all the one loop corrections in the  theory, is an important fundamental concept in quantum field theory. At zero temperature we know that the $n$-point amplitudes (involving the background fields)  at one loop are, in general,  divergent and, consequently, the evaluation of the effective action at $T=0$ needs a regularization. Effective actions can, of course, be evaluated perturbatively. However, a beautiful method due to Schwinger \cite{schwinger}, also known as the proper time formalism, is quite useful in evaluating one loop effective actions at zero temperature with a gauge invariant regularization (in the case of gauge backgrounds). It involves solving dynamical equations in the proper time, which are not always trivial, to determine the effective action. When the dynamical equations can be solved in a closed form, the gauge invariant regularized effective  action can be written in a closed form or at least in an integral representation. This has been profitably used to calculate the imaginary part of the effective action for fermions interacting with a constant background electromagnetic field which describes the decay rate of the vacuum \cite{schwinger}. When the dynamical equations cannot be solved in a closed form, the method due to Schwinger leads to a perturbative determination of the effective action.

In the past couple of decades, there have been several attempts \cite{generalization,frenkel} to generalize the method due to Schwinger to finite temperature \cite{temp,das} and to determine the imaginary part of the effective action leading to conflicting results \cite{generalization}. Here we present an alternative method for determining finite temperature effective actions for fermions interacting with an arbitrary background field. We believe that since the amplitudes at finite temperature are ultraviolet finite unlike those at zero temperature, it is not necessary to generalize the method due to Schwinger to finite temperature. After all, the proper time method was designed to provide a (gauge invariant) regularization which is not necessary at finite temperature. Therefore, we propose a direct method for evaluating finite temperature effective actions based mainly on the general properties of systems at finite temperature. In this connection, we believe that the real time formalism \cite{das} (we use the closed time path formalism due to Schwinger \cite{schwinger1}) is more suited for this purpose. We note that, in general,  the imaginary time formalism (the Matsubara formalism \cite{matsubara})  leads naturally to retarded and advanced amplitudes, but the Feynman (time ordered) amplitudes (beyond the two point function) cannot be consistently generated in this formalism \cite{evans}. On the other hand, the effective action that we are interested in is precisely the one that generates Feynman amplitudes. In contrast to the imaginary time formalism, the effective action, when evaluated properly in the real time formalism, leads naturally not only to the Feynman amplitudes, but also to the retarded and the advanced amplitudes as we will show in examples. Furthermore, as we have emphasized earlier in \cite{das,tor}, the real time calculations can be carried out quite easily in the mixed space where the spatial coordinates have been Fourier transformed as we will describe in the following examples.

Let us consider a system of fermions interacting with an external field which we generically denote by $A$. This can be a scalar or a vector background field and we suppress the Lorentz index (structure) of the background field for simplicity.  If the fermion has a mass $m$, from the definition of the effective action it is straightforward to obtain  
\begin{equation}
\frac{\partial\Gamma_{\rm eff}}{\partial m} = \int \mathrm{d}t \mathrm{d}\mathbf{x}\ {\rm tr}\ S (t,\mathbf{x}; t, \mathbf{x}),\label{propagator0}
\end{equation}
where $S (t,\mathbf{x};t', \mathbf{x}')$ denotes the complete Feynman propagator for the fermion in the presence of the background field and ``tr" stands for the trace over the spinor indices. However, keeping in mind that the fermion may not always have a mass (say, for example, in the Schwinger model \cite{schwinger2}), we use alternatively the fact that the variation of the effective action with respect to the background field leads to the generalized fermion ``propagator" at coincident points (even though we use the same symbol as in \eqref{propagator0}, the exact meaning of $S$ below depends on the nature of the background field as we explain),
\begin{equation}
\frac{\delta \Gamma_{\rm eff}}{\delta A (t,\mathbf{x})} =  g\ {\rm tr}\ S (t, \mathbf{x}; t, \mathbf{x}),\label{propagator1}
\end{equation}
where $g$ denotes the coupling to the background field and we are
suppressing the Lorentz structure of the background field as well as
that of the generalized ``propagator". We note that for a scalar
background, $S$ in \eqref{propagator1} indeed denotes the complete
fermion propagator of the interacting theory at coincident
coordinates. On the other hand, for a gauge field background, the
right hand side in \eqref{propagator1} determines the current density of the theory which is related to the complete fermion propagator of the theory through a Dirac trace involving the Dirac matrix. In either case, we note that it is the fermion propagator that is relevant in \eqref{propagator1} for the evaluation of the effective action. In the mixed space (where the coordinates $\mathbf{x}$ have been Fourier transformed), we can write \eqref{propagator1} as
\begin{equation}
\frac{\delta\Gamma_{\rm eff}}{\delta A (t, -\mathbf{p})} = g\ {\rm tr}\ S (t,t; \mathbf{p}).\label{propagator}
\end{equation} 

Since the effective action is so intimately connected with the fermion propagator and since we are not interested in the zero temperature part of the effective action, our proposal is to determine the complete fermion propagator at finite temperature directly such that 
\begin{enumerate}
\item[(i)] it satisfies the appropriate equations for the complete propagator of the theory, 
\item[(ii)] it satisfies the necessary symmetry properties of the theory such as the Ward identity, 
\item[(iii)] and most importantly, it satisfies the anti-periodicity property associated with a finite temperature fermion propagator \cite{das}. 
\end{enumerate}
In fact, it is the third requirement that is quite important in a direct determination of the propagator. We note that this last condition is missing at zero temperature which makes it difficult to determine the complete propagator (independent of the problem of divergence). When the theory is divergence free (so that it does not need a regularization at zero temperature), this propagator will be the exact fermion propagator of the theory and would lead to the complete effective action including the correct zero temperature part. On the other hand, if the theory needs to be regularized at zero temperature, this propagator will not yield the correct zero temperature effective action, but the finite temperature part of the effective action, which does not need to be regularized, will be determined correctly. We illustrate the method with two examples.

\begin{figure}[ht!]
\begin{center}
\includegraphics{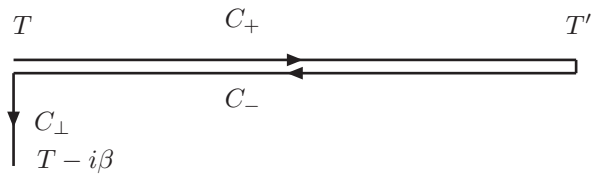}
\end{center}
\caption{The closed time path contour in the complex $t$ plane. Here
  $T\rightarrow -\infty$, while $T'\rightarrow \infty$ and $\beta$
  denotes the inverse temperature (in units of the Boltzmann constant
  $k$) \cite{das}. The two branches labelled by $C_{\pm}$ lead to the
  doubling of the degrees of freedom while the branch $C_{\perp}$ along the imaginary axis decouples from any physical amplitude.}
\label{1}
\end{figure}

Let us start with the $0+1$ dimensional QED described by the Lagrangian
\begin{equation}
L = \overline{\psi} (t) (i\partial_{t} - m - eA (t)) \psi (t),\label{0plus1L}
\end{equation}
where the fermion mass can be thought of as a chemical potential and in $0+1$ dimension, the fermion field as well as  the gauge potential have only a single component. This is a simple model which has been studied exhaustively \cite{0+1, dunne} in connection with large gauge invariance \cite{babu} at finite temperature, but it is also quite useful in clarifying what is involved in our proposal before we generalize it to higher dimensions. As we noted earlier, we use the closed time path formalism where the path in the complex time plane has the form shown in Fig. \ref{1}. In the closed time path formalism (in any real time formalism) \cite{das}, the degrees of freedom need to be doubled and we denote the background fields on the $C_{\pm}$ branches of the contour as $A_{\pm} (t)$ respectively. Since $t$ is the only coordinate on which field variables depend in this theory, there is no need for a mixed space propagator. We note that the complete fermion propagator of the theory (ordered along the contour in Fig. \ref{1}) satisfies the equations
\begin{eqnarray}
& & (i\partial_{t} - m - e A_{c} (t)) S_{c} (t,t') = i\delta_{c} (t-t'),\nonumber\\
& & S_{c}(t,t') (i\overleftarrow{\partial}_{\!\!t'} + m + e A_{c} (t')) = - i\delta_{c} (t-t'),\label{0plus1eqn}
\end{eqnarray}
where the subscript ``$c$" characterizes a function on the contour. On the contour, the step function is defined naturally as \cite{das}  
\begin{equation}
\theta_{c} (t-t') = \left\{\begin{array}{ll}
\theta (t-t') & {\rm if\ both}\  t,t'\in C_{+},\\
\theta (t'-t) & {\rm if\ both}\ t,t' \in C_{-} {\rm or}\ \in C_{\perp},\\
1 & {\rm if}\ t\in C_{-} {\rm and}\ t'\in C_{+},\\
0 & {\rm if}\ t\in C_{+} {\rm and}\ t'\in C_{-}.
\end{array}\right.\label{thetac}
\end{equation}
and we have $\delta_{c} (t-t') = \partial_{t} \theta_{c} (t-t')$.

Equations \eqref{0plus1eqn} can be solved exactly subject to our three requirements leading to the contour ordered propagator of the form
\begin{eqnarray}
S_{c} (t,t') & = & e^{-imt-\frac{ie}{2}\int \mathrm{d}\bar{t}\, {\rm sgn}_{c} (t-\bar{t}) A_{c} (\bar{t})}\nonumber\\
& & \times\left(\theta_{c} (t-t') - n_{\rm F} (m+\frac{ie}{\beta} (a_{+} - a_{-}))\right)\nonumber\\
& & \times e^{imt'+ \frac{ie}{2} \int \mathrm{d}\bar{t}\, {\rm sgn}_{c}(t'-\bar{t}) A_{c} (\bar{t})},\label{0plus1propagator}
\end{eqnarray}
where $n_{\rm F}$ denotes the Fermi distribution function with $\beta$ corresponding to the inverse temperature (in units of the Boltzmann constant $k$) and we have identified 
\begin{equation}
a_{\pm} = \int\limits_{-\infty}^{\infty} \mathrm{d}t\, A_{\pm} (t).\label{0plus1a}
\end{equation}
Furthermore, we have identified
\begin{equation}
{\rm sgn}_{c} (t-t') = \theta_{c} (t-t') - \theta_{c} (t'-t).
\end{equation}
Although the phases in \eqref{0plus1propagator} can be combined to write them in a simpler form in this case, we have chosen to write them in this suggestive form which generalizes naturally to higher dimensions where the propagator will carry spinor indices. When $t,t'$ are restricted to the appropriate branches of the contour, \eqref{0plus1propagator} determines all the components of the full $2\times 2$ matrix propagator of the theory. Furthermore, it can be checked that this propagator satisfies the Lippmann-Schwinger equation (the perturbation expansion) \cite{lippman} for the propagator. However, we do not go into the details of this analysis here, which will be discussed elsewhere. 

The $0+1$ dimensional theory is free from divergences and, therefore, \eqref{0plus1propagator}  represents the complete fermion propagator of the theory in the presence of a background gauge field. We can now take the coincident limit $(t=t')$ and integrate \eqref{propagator1} to obtain the normalized effective action of the theory which has the form
\begin{eqnarray}
\Gamma_{\rm eff} [a_{+},a_{-}] & = & - i \ln \left[\cos \frac{e(a_{+}-a_{-})}{2}\right.\nonumber\\
& & \qquad\left. + i \tanh \frac{\beta m}{2}\sin \frac{e(a_{+}-a_{-})}{2}\right].\label{0plus1effaction}
\end{eqnarray}
This is the complete effective action of the theory which reduces to the well studied action \cite{0+1,dunne} on $C_{+}$ when we set $a_{-}=0$. However, being the complete effective action, \eqref{0plus1effaction} contains all the information about retarded, advanced and other amplitudes as well. For example, let us note from \eqref{0plus1effaction}  that since $\Gamma_{\rm eff} = \Gamma_{\rm eff} [a_{+}-a_{-}]$, the retarded $n$-point amplitude of the theory can be shown to vanish (for a definition of retarded amplitudes, see \cite{retarded}), namely,
\begin{eqnarray}
\Gamma_{R}^{(n)} & = & \sum_{m=0}^{n-1}\left(^{n-1}\!C_{m}\,\frac{\mathrm{d}^{n-1-m}}{\mathrm{d}a_{+}^{n-1-m}} \frac{\mathrm{d}^{m}}{\mathrm{d} a_{-}^{m}}\right)\frac{\mathrm{d}\Gamma_{\rm eff}[a_{+}-a_{-}]}{\mathrm{d}a_{+}}\bigg|\nonumber\\
& = & (1-1)^{n-1}\, \frac{\mathrm{d}^{n}\Gamma_{\rm eff}[a_{+}-a_{-}]}{\mathrm{d} a_{+}^{n}}\bigg|  = 0, \quad n\geq 2,\label{vanishingR}
\end{eqnarray}
where the restriction stands for setting all the background fields to zero. Therefore, all the retarded (advanced) amplitudes vanish in this theory. (The one point amplitude is by definition a Feynman amplitude.) 

With this brief derivation of the complete effective action in the $0+1$ dimensional theory, let us next consider the fermion sector of the Schwinger model \cite{schwinger2} or massless QED in $1+1$ dimensions described by the Lagrangian density
\begin{equation}
{\cal L} = \overline{\psi} (t,x) \gamma^{\mu} (i\partial_{\mu} - eA_{\mu} (t,x)) \psi  (t,x).\label{schwingerL}
\end{equation}
At zero temperature, this model is soluble and describes free massive photons. The effective action for this model (for an arbitrary gauge background) has also been studied perturbatively at finite temperature \cite{adilson} even in the presence of a chemical potential \cite{silvana}. Here we will derive the closed form expression of the finite temperature effective action following our method. We note here that the two point function in the Schwinger model needs to be regularized at zero temperature and, consequently, the zero temperature part of the effective action following from our propagator will not coincide with the regularized zero temperature effective action. However, our interest is in the finite temperature part of the effective action which is free from ultraviolet divergences. For completeness we note that the simple point-splitting regularization of the fermion propagator is sufficient to regularize the theory and can be carried out even in our method. However, we will not do this here since our main interest is in the finite temperature part of the effective action.

The theory \eqref{schwingerL} is best studied in the natural basis of right handed and left handed fermion fields (although everything that we say can be carried out covariantly as well as in the presence of a chemical potential). Defining \cite{dasbook,glf}
\begin{eqnarray}
\psi_{\rm R} & = &  \frac{1}{2} (\mathbbm{1}+\gamma_{5})\psi, \quad \psi_{\rm L} = \frac{1}{2} (\mathbbm{1} - \gamma_{5}) \psi,\nonumber\\
x^{\pm} & = & \frac{x^{0}\pm x^{1}}{2}, \quad p_{\pm} = p_{0}\pm p_{1},\quad \partial_{\pm} = \partial_{0}\pm \partial_{1},\nonumber\\ 
A_{\pm} & = & A_{0} \pm A_{1},\label{light-conevariables}
\end{eqnarray}
the Lagrangian density \eqref{schwingerL} naturally decomposes into two decoupled sectors described by 
\begin{equation}
{\cal L} = \psi_{\rm R}^{\dagger} (i\partial_{+} - eA_{+})\psi_{\rm R} + \psi_{\rm L}^{\dagger} (i\partial_{-} - eA_{-})\psi_{\rm L},\label{rlsectors}
\end{equation}
where $\psi_{\rm R}, \psi_{\rm L}$ denote only the component spinor fields (no spinor index left any more). While the zero temperature regularization mixes the two sectors through the two point function (anomaly), at finite temperature we do not have divergences and we do not expect the two sectors to mix. Therefore, we can study the finite temperature effective action in each of the two sectors separately.

Let us consider the theory only in the sector of the right handed
fermions in \eqref{rlsectors}. This is very much like the $0+1$
dimensional theory. However, there is one essential difference which
makes the derivation much more difficult, namely, the field variables
depend on two coordinates $(t,x)$ or equivalently on
$(x^{+},x^{-})$. We would like to emphasize here that although we use
the light-cone coordinates for simplicity, the theory is still
quantized on the equal-time surface and the propagator is defined
through the time ordered Green's function (namely, we do not use the
statistical mechanics of the light-front \cite{lightfront}). As we
mentioned earlier, the finite temperature derivations become a lot
simpler in the mixed space. Thus, Fourier transforming the $x^{-}$
coordinate, the action for the right handed fermions takes the form
(the conjugate variables to $x^{-}$ should be written as $p_{-},k_{-}$, which we write as $p,k$ for simplicity)
\begin{eqnarray}
S_{\rm R} & = & 2 \int \mathrm{d}x^{+} \frac{\mathrm{d}p}{2\pi}\,\psi_{\rm R}^{\dagger} (x^{+},-p)\bigg[i\partial_{+} \psi_{\rm R} (x^{+},p)\nonumber\\
 & & \quad  -e\int \frac{\mathrm{d}k}{2\pi}\, A_{+} (x^{+},p-k) \psi_{\rm R} (x^{+},k)\bigg].\label{rtreeaction}
\end{eqnarray}
As a result, we recognize that the equations for the propagator will involve a convolution. They are best described by introducing the following operator notations for the propagator as well as the gauge potential
\begin{eqnarray}
S (x^{+},x'^{+};p,k) & = & \langle p|\hat{S} (x^{+},x'^{+})|k\rangle,\nonumber\\ 
A_{+} (x^{+},p-k) & =  & \langle p|\hat{A}_{+} (x^{+})|k\rangle,\label{operators}
\end{eqnarray}
so that the equations for the propagator ordered along the contour take the (operator) forms (see also \eqref{0plus1eqn})
\begin{eqnarray}
&\!\!\!&\!\!\!\!\!\!\!(i\partial_{+} - e \hat{A}_{+c} (x^{+})) \hat{S}_{c} (x^{+},x'^{+}) = \frac{i}{2}\,\delta_{c} (x^{+}-x'^{+}),\nonumber\\
&\!\!\!&\!\!\!\!\!\!\!\!\! \hat{S}_{c} (x^{+},x'^{+}) (i\overleftarrow{\partial}'_{\!\!+} + e \hat{A}_{+c} (x'^{+})) = - \frac{i}{2}\, \delta_{c} (x^{+}-x'^{+}).\label{1plus1eqn}
\end{eqnarray} 
We note from \eqref{propagator} and \eqref{rtreeaction} that, in the present case,  we can identify
\begin{equation}
S_{c} (x^{+},x'^{+};p) = \int \frac{\mathrm{d}k}{2\pi}\, \langle k+p|\hat{S}_{c} (x^{+},x'^{+})|k\rangle.
\end{equation}

The solution to \eqref{1plus1eqn} satisfying the Ward identity as well as the appropriate anti-periodicity condition can be determined to have the form
\begin{eqnarray}
\hat{S}_{c} (x^{+},x'^{+})  & = & \frac{1}{4}\, e^{-\frac{ie}{2} \int \mathrm{d}\bar{x}^{+}\,{\rm sgn}_{c} (x^{+}-\bar{x}^{+}) \hat{A}_{+c} (\bar{x}^{+})}\nonumber\\
& & \times \big({\rm sgn}_{c} (x^{+}-x'^{+}) + 1 - 2(\hat{\cal O}_{+} + 1)^{-1}\big)\nonumber\\
& & \times e^{\frac{ie}{2} \int \mathrm{d}\bar{x}^{+}\,{\rm sgn}_{c} (x'^{+}-\bar{x}^{+}) \hat{A}_{+c} (\bar{x}^{+})},\label{1plus1propagator}
 \end{eqnarray}
 where
 \begin{eqnarray}
& & {\rm sgn}_{c} (x^{+}-x'^{+}) = \theta_{c} (x^{+}-x'^{+}) - \theta_{c}(x'^{+}-x^{+}),\nonumber\\
& & \hat{\cal O}_{+} =  e^{\frac{ie (\hat{a}_{+ (+)}-\hat{a}_{+ (-)})}{2}}\,e^{\frac{\beta K}{2}}\,e^{\frac{ie (\hat{a}_{+(+)}-\hat{a}_{+(-)})}{2}},\label{operatorO}
 \end{eqnarray}
with $K$ denoting the momentum operator and ($(\pm)$ with the parenthesis denote the thermal indices while $+$ without the parenthesis represents the light-cone component of the background field)
 \begin{equation}
 \hat{a}_{+ (\pm)} = \int\limits_{-\infty}^{\infty} \mathrm{d}x^{+}\,\hat{A}_{+(\pm)} (x^{+}).
 \end{equation}
 It can be checked that  this complete propagator satisfies the Lippmannn-Schwinger equation. In fact, setting $x^{+}=x'^{+}$ and using \eqref{propagator} we can integrate \eqref{1plus1propagator} to obtain the normalized effective action in the right handed sector. The thermal part of the effective action is contained in the simple form  
\begin{equation}
\Gamma_{\rm R,\, eff} =  -\frac{i}{2} \int  \frac{\mathrm{d}k}{2\pi}\,\langle k|\ln \big(1 + \frac{ie}{2}\,\hat{N} \hat{a}_{+}\big)|k\rangle,\label{raction}
\end{equation}
with  
\begin{equation}
\hat{N} = 1-2n_{\rm F} (\frac{K}{2}),\quad \hat{a}_{+} = \hat{a}_{+(+)} - \hat{a}_{+(-)}.
\end{equation} 
This effective action has the right (delta function) structure that had already been observed in the perturbative calculation in the right handed sector \cite{adilson} which is a consequence of the Ward identity in the theory. In fact, the expansion of this effective action on $C_{+}$ (namely, setting $A_{+(-)}=0$) agrees order by order with the perturbative result. The thermal part of the effective action for the left handed sector is similarly given by 
\begin{equation}
\Gamma_{\rm L,\, eff}  = -\frac{i}{2} \int \frac{\mathrm{d}k}{2\pi}\,\langle k|\ln \big(1 + \frac{ie}{2}\,\hat{N} \hat{a}_{-}\big)|k\rangle,\label{laction}
\end{equation}
where $k$ should be understood as the conjugate variable to $x^{+}$
(which should be written as $k_{+}$) and  
 we have identified
\begin{eqnarray}
\hat{a}_{-} & = &\hat{a}_{-(+)}-\hat{a}_{-(-)},\nonumber\\
\hat{a}_{-(\pm)} & = & \int\limits_{-\infty}^{\infty} \mathrm{d}x^{-}\,\hat{A}_{-(\pm)} (x^{-}).
\end{eqnarray}

Once again, this effective action has the right (delta function) structure as in the perturbative calculation and the thermal part agrees with the perturbative result \cite{adilson} order by order when restricted to $C_{+}$. The finite temperature effective action for the  $1+1$ dimensional fermion interacting with an arbitrary Abelian gauge background can, therefore, be written as
\begin{equation}
\Gamma_{\rm eff} = \Gamma_{\rm R, eff} + \Gamma_{\rm L, eff}\label{complete}
\end{equation}
and the thermal part of \eqref{complete} leads to the correct perturbative result order by order \cite{adilson} on the branch $C_{+}$. However, since \eqref{complete} represents the complete effective action and since it is a functional of $(\hat{a}_{\pm (+)} - \hat{a}_{\pm (-)})$ (see \eqref{raction}-\eqref{laction}), it can be checked as in \eqref{vanishingR} that all the retarded (advanced) amplitudes vanish in this theory. This should be contrasted with the fact that this had been verified explicitly only up to the 4-point function in perturbation theory  \cite{retarded}.
 
In summary, we have proposed an alternative method for determining the finite temperature effective action for fermions interacting with an arbitrary background field. This is done by determining the complete fermion propagator (in the closed time path formalism) directly by using the anti-periodicity condition appropriate at finite temperature. We have illustrated how our proposal works with the examples of the $0+1$ dimensional QED as well as the Schwinger model.  A longer version of the results with more details of the calculations as well as other aspects of this analysis will be reported  separately.

\bigskip

\noindent{\bf Acknowledgments}
\medskip

This work was supported in part  by US DOE Grant number DE-FG 02-91ER40685,  by CNPq and FAPESP (Brazil).

%\bibliographystyle{prsty}
%\bibliography{all_new}

\begin{thebibliography}{10}

\bibitem{schwinger} J. Schwinger, Phys. Rev. {\bf 82}, 664 (1951).

\bibitem{generalization} W. Dittrich, Phys. Rev. {\bf D19}, 23 (1978); P. H. Cox, and W. S. Hellman, Ann. Phys. {\bf 154}, 211 (1984); M. Loewe and J. C. Rojas, Phys. Rev. {\bf D46}, 2689 (1992); P. Elmfors, D. Persson and B.-S. Skagerstam, Phys. Rev. Lett. {\bf 71}, 480 (1993); P. Elmfors and B.-S. Skagerstam, Phys. Lett. {\bf B348}, 141 (1995); H. Gies, Phys. Rev. {\bf D60}, 105002 (1999); S. P. Gavrilov and D. M. Gitman, Phys. Rev. {\bf D78}, 045017 (2008).

\bibitem{frenkel} A. Das and J. Frenkel, Phys. Rev. {\bf D75}, 025021 (2007).

\bibitem{temp} J. Kapusta, {\em Finite Temperature Field Theory}, Cambridge University Press, Cambridge, England (1989); M. Le Bellac, {\em Thermal Field Theory}, Cambridge University Press, Cambridge, England (1996).

\bibitem{das} A. Das, {\em Finite Temperature Field Theory}, World Scientific, Singapore (1997).

\bibitem{schwinger1} J. Schwinger, {\em Lecture Notes of Brandeis Summer Institute in Theoretical Physics} (1960); J. Schwinger, J. Math. Phys. {\bf 2}, 407 (1961); P. M. Bakshi and K. T. Mahanthappa, J. Math. Phys. {\bf 4}, 1 (1963); L. V. Keldysh, Sov. Phys. JETP {\bf 20}, 1018 (1965).

\bibitem{matsubara} T. Matsubara, Prog. Theor. Phys. {\bf 14}, 351 (1954).

\bibitem{evans} T. S. Evans, Nucl. Phys. {\bf B374}, 340 (1992).

\bibitem{tor}  F. T. Brandt, A. Das, O. Espinosa, J. Frenkel and S. Perez, Phys. Rev. {\bf D72}, 085006 (2005); {\em ibid} {\bf D73}, 065010 (2006); {\em ibid} {\bf D73}, 067702 (2006).

\bibitem{schwinger2} J. Schwinger, Phys. Rev. {\bf 128}, 2425 (1962).

\bibitem{0+1} G. Dunne, K. Lee and C. Lu, Phys. Rev. Lett. {\bf 78}, 3434 (1997)

\bibitem{dunne} A. Das and G. Dunne, Phys. Rev. {\bf D57}, 5023 (1998); J. Barcelos-Neto and A. Das, Phys. Rev. {\bf D58}, 085022 (1998).

\bibitem{babu} K. S. Babu, A. Das and P. Panigrahi, Phys. Rev. {\bf D36}, 3725 (1987).

\bibitem{lippman} B. A. Lippmann and J. Schwinger, Phys. Rev. {\bf 79}, 469 (1950).

\bibitem{retarded} F. T. Brandt, A. Das, J. Frenkel and A. J. da Silva, Phys. Rev. {\bf D59}, 065004 (1999); F. T. Brandt, A. Das and J. Frenkel, Phys. Rev. {\bf D60}, 105008 (1999).

\bibitem{adilson} A. Das and A. J. da Silva, Phys. Rev. {\bf D59}, 105011 (1999).

\bibitem{silvana}  S. Maciel and S. Perez, Phys. Rev. {\bf D78}, 065005 (2008).

\bibitem{dasbook}
For quantization of massless fermion fields, see, for example, {\em Lectures on Quantum Field Theory}, A. Das, World Scientific Publishing, Singapore (2008).

\bibitem{glf} For a general discussion of transformations between different coordinate systems, see, for example, A. Das and S. Perez, Phys. Rev. {\bf D70}, 065006 (2004).

\bibitem{lightfront} V. S. Alves, A. Das and S. Perez, Phys. Rev. {\bf D66}, 125008 (2002).


\end{thebibliography}
\end{document}